\documentclass[runningheads]{llncs}
\usepackage{graphicx}
\usepackage{url}

\usepackage{indentfirst}
\usepackage{marvosym}
\newcommand{\tabincell}[2]{\begin{tabular}{@{}#1@{}}#2\end{tabular}}
\begin{document}
\title{Edge AIBench: Towards Comprehensive End-to-end Edge Computing Benchmarking}
%
%
\author{Tianshu Hao\inst{1,2} \and
Yunyou Huang \inst{1,2} \and
Xu Wen \inst{1,2} \and
Wanling Gao \inst{1,3} \and
Fan Zhang \inst{1} \and
Chen Zheng \inst{1,3} \and
Lei Wang \inst{1,3} \and
Hainan Ye \inst{3,4} \and
Kai Hwang \inst{5} \and
Zujie Ren \inst{6} \and
Jianfeng Zhan\inst{1,2,3,}$^\ast$ \thanks{Jianfeng Zhan is the corresponding author.}}

\authorrunning{Tianshu H, et al.}
%
\institute{State Key Laboratory of Computer Architecture, Institute of Computing Technology, Chinese Academy of Sciences \and
University of Chinese Academy of Sciences \and
BenchCouncil (International Open Benchmarking Council) \and
Beijing Academy of Frontier Sciences and Technology \and
Chinese University of Hongkong at Shenzhen \and
Zhejiang Lab
\\$^\ast$ \email{zhanjianfeng@ict.ac.cn}
}

\maketitle              
\begin{abstract}
In edge computing scenarios, the distribution of data and collaboration of workloads on different layers are serious concerns for performance, privacy, and security issues. So for edge computing benchmarking, we must take an end-to-end view, considering all three layers: client-side devices, edge computing layer, and cloud servers. Unfortunately, the previous work ignores this most important point. This paper presents the BenchCouncil's coordinated effort on edge AI benchmarks, named Edge AIBench. In total, Edge AIBench models four typical application scenarios: ICU Patient Monitor, Surveillance Camera, Smart Home, and Autonomous Vehicle with the focus on data distribution and workload collaboration on three layers. Edge AIBench is a part of the open-source AIBench project, publicly available from\url{http://www.benchcouncil.org/AIBench/index.html}. We also build an edge computing testbed with a federated learning framework to resolve performance, privacy, and security issues.

\keywords{Edge Computing \and AI Benchmarks \and Testbed \and Federated Learning.}
\end{abstract}
\section{Introduction}
\begin{figure}[htbp]
\centering
\includegraphics[width=8cm]{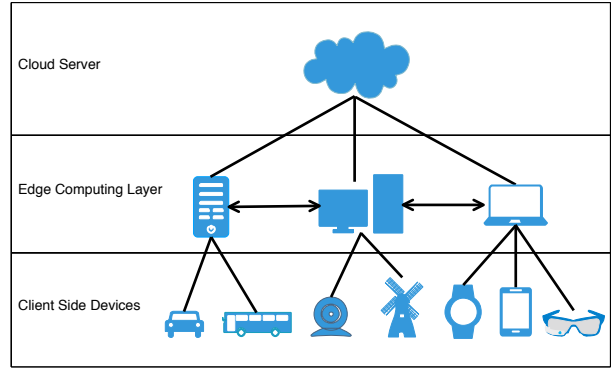}
\caption{A General Edge Computing Framework.}
\label{edge_computing}
\end{figure}

Cloud computing is a mature model to share computing resources by providing network access to users~\cite{cloud_computing}. In cloud computing models,  users communicate with the data center to get hardware, software and other computing resources and store data. However, In recent years, the number of client-side devices (e.g. smart devices and monitors) grows rapidly.  IoT Analytics~\cite{iot_analytics} has reported the number of connected devices reached 17 billion in 2018 and Gartner says the IoT devices will install 26 billion units by 2020~\cite{Gartner}. These client-side devices produce a large amount of data to process. The overhead of data transmission and data encryption among devices and data centers becomes significant bottlenecks for many IoT scenarios, and hence it raises a daunting challenge for throughput, latency, and security guarantee.
\\ \indent Edge computing emerges as a promising technical framework to overcome the challenges in cloud computing. The edge computing framework adds a new layer,  named the edge computing layer, on the basis of the traditional cloud computing framework. In the edge computing framework, only the real-time data processing is transferred to the edge computing layer, while other complicated data processing is still executed on the cloud server. Figure~\ref{edge_computing} shows a general edge computing framework, which includes three layers: cloud server, edge computing layer, and client-side devices.
\\ \indent In the edge computing scenarios, the distribution of data and collaboration of workloads on different layers are serious concerns for performance, security, and privacy issues. So for benchmarking, designing, and implementing edge computing systems or applications, we shall take an end-to-end view, considering all three layers. Unfortunately, the previous work, especially the previous benchmarking efforts~\cite{MLPerf,EEMBC,AIBenchmark,edgebench} ignore this most important point.
\\ \indent In edge computing scenarios, AI techniques are widely used to augment device, edge and cloud intelligence, and they are most demanding in terms of computing power, data storage, and network. Typical application scenarios include smart city, smart home, autonomous vehicle, surveillance camera, smart medical, wearable devices and so on. These scenarios are complicated because of different kinds of client-side devices, a large quantity of heterogeneous data, privacy and security issues. Most of these scenarios have a high requirement for latency and network bandwidth. However, edge computing is in the initial stage and doesn't have a uniform standard for these scenarios. Therefore, a comprehensive end-to-end edge computing benchmark suite is needed to measure and optimize the systems and applications. 
\\ \indent Meanwhile, edge computing is still in the initial stage with a lack of testbed. Because of the privacy issue, there is no incentive to share data. Because of the complexity, there is no end-to-end application scenario to validate the architectures, systems, or specific algorithms in certain settings.
\\ \indent Above all, it's necessary to develop a benchmark suite and testbed for edge computing. This paper reports the  BenchCouncil's coordinated effort on edge AI benchmarks, named Edge AIBench. Edge AIBench is a part of the open-source AIBench project, which is publicly available from\url{http://www.benchcouncil.org/AIBench/index.html}. Edge AIBench includes four typical application scenarios: ICU Patient Monitor, Surveillance Camera, Smart Home, and Autonomous Vehicle, which consider the complexity of all edge computing AI scenarios. Coordinated by BenchCouncil (http://www.benchcouncil.org), we are also building an edge computing testbed with a federated learning framework to resolve security and privacy issue, which can be accessed from\url{http://www.benchcouncil.org/testbed/index.php}. BenchCouncil also release datacenter AI benchmarks~\cite{AI}, HPC AI benchmarks~\cite{HPCAI}, and IoT AI benchmarks~\cite{IoTAI}, publicly available from\url{http://www.benchcouncil.org/AIBench/index.html}.

\section{Related Work}
Since the edge computing AI applications have become more and more popular these years, benchmarks are needed to measure and optimize the systems and applications. There are several related benchmark suites. We summarize the state-of-the-art and state-of-the-practice work on edge AI benchmarking.
\\ \indent MLPerf~\cite{MLPerf} is a benchmark suite focusing on measuring Machine Learning(ML) performance. It provides the edge inference benchmarks, including eight ML tasks: image classification, object detection and so on~\cite{MLPerf_tutorial}. But this benchmark suite just evaluates the edge computing layer with the lack of an end-to-end view.
\\ \indent
EEMBC~\cite{EEMBC} develops an ML benchmark suite, named MLMark on embedded edge computing platforms. MLMark includes four AI applications: image classification, object detection, language translation, and speech recognition. However, only the licensees and members of EEMBC have the right to access these benchmarks and this benchmark suite is still in ``beta" state now.
\\ \indent EdgeBench~\cite{edgebench} compares two edge computing platforms--Amazon AWS Greengrass and Microsoft Azure IoT Edge. And it includes two AI applications: speech-to-text and image recognition. 
EdgeBench fails to provide an end-to-end application benchmarking framework.
\\ \indent AI Benchmark~\cite{AIBenchmark} is a benchmark suite for AI applications on smartphones, and it includes nine AI applications. It's an IoT benchmark suite and only focuses on the client-side devices (smartphones)' performance.
\\ \indent Table~\ref{comparison} compares the state-of-the-art and state-of-the-practice edge computing AI benchmarks. It shows many of them only focus on the edge computing layer instead of the whole edge computing framework. Our benchmark suite Edge AIBench provides an end-to-end application benchmarking framework, including train, validate and inference stages. Moreover, Edge AIBench includes four typical edge computing AI scenarios and measures the whole three-layer edge computing framework.
\begin{table}
\label{comparison}
\begin{center}
\caption{Comparison among Edge Computing AI benchmarks}

\begin{tabular}{|p{2cm}|p{3cm}|p{1.8cm}|p{1.8cm}|p{1.8cm}|p{1cm}|}

\hline
Benchmark Name &End-to-end Application Scenarios &Components on Cloud Server &Components on Edge Computing Layer &Components on Client-side Devices& Open-Source
\\
\hline
Edge AIBench&\tabincell{l}{ICU Patient Monitor\\Surveillance Camera\\Smart Home\\ Autonomous Vehicle}&$\surd$&$\surd$&$\surd$&$\surd$
\\
\hline
MLPerf&N/A&${\times}$&$\surd$&${\times}$&$\surd$
\\
\hline
EEMBC  MLMark&Not Clear&Not Clear&Not Clear&Not Clear&${\times}$
\\
\hline
EdgeBench&N/A&$\surd$&$\surd$&$\surd$&$\surd$
\\
\hline
AI Benchmark&N/A&${\times}$&${\times}$&$\surd$&$\surd$
\\
\hline

\end{tabular}
\end{center}
\end{table}

\section{The summary of Edge AIBench}
Edge AIBench includes four typical scenarios: Intensive care unit(ICU) patient monitor, surveillance camera, smart home, and autonomous vehicle. These four AI scenarios can present the complexity of edge computing AI scenarios from different perspectives.

\subsection{ICU Patient Monitor}
ICU is the treatment place for critical patients. Therefore immediacy is significant for ICU patient monitor scenario to notify doctors of the patients' status as soon as possible. The dataset we use is MIMIC-III~\cite{mimic}. MIMIC-III provides many kinds of patients data such as vital signs, fluid balance and so on. Moreover, we choose heart failure prediction~\cite{retain} and endpoint prediction~\cite{endpoint} as the AI benchmarks.
\\ \indent Heart failure prediction uses the MIMIC-III dataset and a two-level neural attention model. It collects the patients' data on the virtual client-side devices, trains on the cloud server (the data will be sent from the edge) and predicts the heart failure on the edge computing layer.
\\ \indent Endpoint prediction benchmark also uses the MIMIC-III dataset, and it uses an LSTM model. This benchmark collects patients' data on the virtual patient device generator and then transmit it to the edge to make the inference. Then the data will be sent to the cloud server to do more training.
\subsection{Surveillance Camera}
There are many surveillance cameras all over the world nowadays, and these cameras will produce a large quantity of video data at all times. If we transmit all of the data to cloud servers, the network transmission bandwidth will be very high. Therefore, this scenario focus on edge data preprocesses and data compression.
\\ \indent We choose the person re-identification application as the component benchmark. It collects data from the virtual camera devices and pre-process and infer these video data on the edge computing layer. Then the edge computing layer will send the compressed data to the cloud server. Moreover, the decompression and training process are on the cloud server.
\subsection{Smart Home}
Smart home includes a lot of smart home devices such as automatic controller, alarm system, audio equipment and so on. Thus, the uniqueness of the smart home includes different kinds of edge devices and heterogeneous data. We will choose two AI applications as the component benchmarks: speech recognition and face recognition. These two components have heterogeneous data and different collecting devices. These two component benchmarks both collect data on the client side devices(e.g. camera and smartphone), infer on the edge computing layer and train on the cloud server.
\\ \indent Speech recognition uses the DeepSpeech2~\cite{deepspeech2} model and the LibriSpeech dataset~\cite{librispeech}.
\\ \indent Face recognition uses the FaceNet~\cite{facenet} model and uses the LFW(Labeled Faces in the Wild)~\cite{lfw} dataset.
\subsection{Autonomous Vehicle}
The uniqueness of the autonomous vehicle scenario is that the high demand for validity. That is to say, it takes absolute correct action even without human intervention. This feature represents the demand of some edge computing AI scenarios. The automatic control system will analyze the current road conditions and make a corresponding reaction at once. We will choose the road sign recognition as the component benchmark.
\\ \indent The road sign recognition will collect the road signs data from the camera, train these data on the cloud and infer on the edge computing layer.
\\ \indent Table~\ref{component} shows the component benchmarks of Edge AIBench. Edge AIBench provides an end-to-end application benchmarking, consisting of train, inference, data collection and other parts using a general three-layer edge computing framework.

\begin{table}
\begin{center}
\caption{The Summary of Edge AIBench}
\begin{tabular}{|p{2cm}|p{3cm}|p{2.5cm}|p{2.5cm}|p{1.5cm}|}
\hline
End-to-end Application Scenarios &AI Component Benchmarks&Cloud Server&Edge Computing Layer&Client Side Device\\
\hline
ICU Patient Monitor& Heart Failure Prediction&Train&\tabincell{l}{Infer\\Send Alarm}&Generate Data \\
\hline
ICU Patient Monitor& Endpoint Prediction &Train&Infer&Generate Data\\
\hline
Surveillance Camera&Person Re-Identification&\tabincell{l}{Decompress Data\\Train}&\tabincell{l}{Compress Data\\Infer}&Generate Data\\
\hline
Smart Home&Speech Recognition&Train&Infer&Generate Data \\
\hline
Smart Home &Face Recognition&Train&Infer&Generate Data\\
\hline
Autonomous Vehicle&Road Sign Recognition&Train&Infer&Generate Data\\
\hline
\end{tabular}
\label{component}
\end{center}
\end{table}

\subsection{A Federated Learning Framework Testbed}
\begin{figure}[htb]
\centering
\includegraphics[width=8cm]{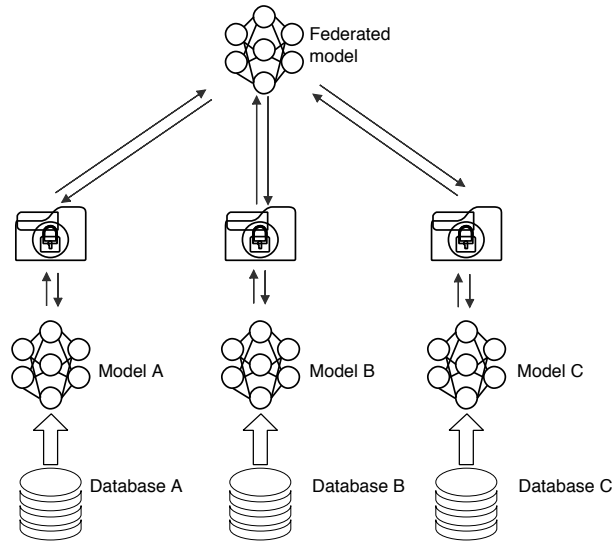}
\caption{An Edge Computing AI Testbed with a Federated Learning Framework}
\label{testbed_framework}
\end{figure}

We have developed an edge computing AI testbed to provide support for researchers and common users, which is publicly available from \url{http://www.benchcouncil.org/testbed.html}. Security and privacy issues become significant focuses in the age of big data, as well as edge computing. Federated learning is a distributed collaborative machine learning technology whose main target is to preserve the privacy~\cite{federated_learning}. Our testbed system will combine the federated learning framework.
\\ \indent At present, we are implementing the ICU scenario on the testbed. We are developing a ``virtual patient" data generator and a federated machine learning training model. Doctors can train the model on the local server and transmit the encrypting parameter to the cloud server. Then the cloud server computes the overall parameter on the basis of these encrypting parameter from different hospitals. After all, the cloud server will send the overall parameter to the local server and the local server will decrypt it to update their models. Figure~\ref{testbed_framework} shows our federated learning testbed framework.

\section{Conclusion}
This paper presents an edge computing AI benchmark, named Edge AIBench, which consists of four end-to-end application benchmarking framework and six component benchmarks. These scenarios we choose can present the complexity of edge computing scenarios from different perspectives. Also, we build an edge computing AI testbed with a federated learning framework.

\section{Acknowledgment}
This work is supported by the Standardization Research Project of Chinese Academy of Sciences No.BZ201800001.
%
%
%
%

\end{document}